\documentclass[trackchanges]{aastex63}

\pdfoutput=1

\received{Jul 07, 2020}
\revised{Aug 27, 2020}
\accepted{Aug 28, 2020}
\submitjournal{ApJL}

\shorttitle{Direct evidence of shock self-reformation}
\shortauthors{Yang et al.}
\begin{document}

\title{MMS direct observations of kinetic-scale shock self-reformation}

\correspondingauthor{Ying D. Liu}
\email{liuxying@swl.ac.cn}

\author[0000-0002-1509-1529]{Zhongwei Yang}
\affiliation{State Key Laboratory of Space Weather, National Space Science Center, Chinese Academy of Sciences, Beijing, 100190, People's Republic of China {\rm{\color{blue}(zwyang1984@gmail.com, liuxying@swl.ac.cn)}}}
\affiliation{Key Laboratory of Polar Science, MNR, Polar Research Institute of China, Shanghai 200136, People's Republic of China}
\affiliation{Key Laboratory of Earth and Planetary Physics, Institute of Geology and Geophysics, Chinese Academy of Sciences, Beijing, 100029, People's Republic of China}

\author[0000-0002-3483-5909]{Ying D. Liu$^{\dag}$}
\affiliation{State Key Laboratory of Space Weather, National Space Science Center, Chinese Academy of Sciences, Beijing, 100190, People's Republic of China {\rm{\color{blue}(zwyang1984@gmail.com, liuxying@swl.ac.cn)}}}
\affiliation{University of Chinese Academy of Sciences, Beijing, 100049, People's Republic of China}

\author[0000-0001-7714-1870]{Andreas Johlander}
\affiliation{University of Helsinki, Helsinki, Finland}

\author[0000-0001-5580-5621]{George K. Parks}
\affiliation{Space Sciences Laboratory, University of California, Berkeley, CA, USA}

\author[0000-0001-6807-8494]{Benoit Lavraud}
\affiliation{Institut de Recherche en Astrophysique et Plan\'{e}tologie, Universit\'{e} de Toulouse, Toulouse, France}

\author[0000-0002-7737-0339]{Ensang Lee}
\affiliation{Kyung Hee University, Yongin, Korea, Republic of (South)}

\author[0000-0001-6271-0110]{Wolfgang Baumjohann}
\affiliation{Space Research Institute, Austrian Academy of Sciences, Graz, Austria}

\author{Rui Wang}
\affiliation{State Key Laboratory of Space Weather, National Space Science Center, Chinese Academy of Sciences, Beijing, 100190, People's Republic of China {\rm{\color{blue}(zwyang1984@gmail.com, liuxying@swl.ac.cn)}}}

\author[0000-0003-0452-8403]{James L. Burch}
\affiliation{Southwest Research Institute, San Antonio, TX, USA}



\begin{abstract}

Studies of shocks have long suggested that a shock can undergo cyclically self-reformation in a time scale of ion cyclotron period. This process has been proposed as a primary mechanism for energy dissipation and energetic particle acceleration at shocks. Unambiguous observational evidence, however, has remained elusive. Here, we report direct observations for the self-reformation process of a collisionless, high Mach number, quasi-perpendicular shock using MMS measurements. We find that reflected ions by the old shock ramp form a clear phase-space vortex, which gives rise to a new ramp. The new ramp observed by MMS2 has not yet developed to a mature stage during the self-reformation, and is not strong enough to reflect incident ions. Consequently, these ions are only slightly slowed down and show a flat velocity profile from the new ramp all the way to the old one. The present results provide direct evidence for shock self-reformation, and also shed light on energy dissipation and energetic particle acceleration at collisionless shocks throughout the universe.

\end{abstract}


% Use the new Unified Astronomy Thesaurus concepts
%\keywords{methods: numerical --- shock waves --- Plasmas --- instabilities --- Sun: coronal mass ejections (CMEs)}


\section{Introduction}

Collisionless shocks are a fundamental phenomenon in the inner heliospheric \citep{Bale2005,Liu2019,Wilson2020}, outer heliospheric \citep{Richardson2008,Zank2015a,Zank2015b}, astrophysical \citep{Plotnikov2019,Guo2012}, and laboratory plasmas \citep{Schaeffer2019}. They are of great interest because they play key roles in energy dissipation and energetic particle acceleration throughout the universe \citep{Blandford1987,Lembege2004,Burgess2005,Wilson2019}. A fraction of the incoming ions can be reflected at high Mach number (super-critical) shocks \citep{Lembege1987,Treumann2009,Sundberg2017}. Numerical simulations suggest that the reflected ions accumulate in front of the shock ramp, which results in a growing foot in the magnetic field. Under certain conditions, the reflected ions and decelerated incident ions form a vortex in the phase-space, which enhances the foot into a new shock ramp \citep{Hada2003,Scholer2003}. As time goes on, the new ramp replaces the old one. This entire process is repeated periodically and named as shock front self-reformation \citep{Lembege1992,Chapman2005,Umeda2006,Caprioli2015}.

Observational studies of shock front nonstationarity have been performed based on only magnetic field measurements \citep{Horbury2001, Mazelle2010}. They point out that more detailed analysis, in particular those combining data from field and particle instruments, will greatly improve the understanding of the collisionless shock phenomenon. Cluster satellites observed quasi-cyclically evolving events of reflected ions at the foot region of Earth's bow shock \citep{Lobzin2007}. However, features of ion velocity distribution functions (VDF) at the reforming ramps cannot be discriminated because the cadence of their ion measurements is only 4 s. Hence the ion phase-space vortex in small scales responsible for the new ramp formation and corresponding shock microstructures cannot be identified unambiguously. \citet{Johlander2016} and \citet{Gingell2017} argue that shock front rippling could be another source of shock non-stationarity. Identification of the shock self-reformation is difficult because self-reformation and rippling are competing and mixed \citep{Lembege2009,Umeda2018,Hao2017,Yang2018}. The following conditions are required to identify shock self-reformation: (1) multi-spacecraft observations; (2) appropriate spacecraft separation and configuration; and (3) high cadence measurements of particle velocity distributions and magnetic fields.

Here, we present direct in situ observations of full particle dynamics and electromagnetic microstructures associated with the self-reformation of Earth's bow shock using high-resolution measurements from the MMS mission \citep{Burch2016}. In order to probe the evolving microstructures with multi-point measurements, the spacecraft separations are required to be as small as 10 km. Three-dimensional ion and electron velocity distributions are provided by the FPI (Fast Plasma Investigation) instrument at a time resolution of 150 ms and 30 ms, respectivley \citep{Pollock2016}. These high cadence, multi-spacecraft plasma measurements allow detailed kinetic scale studies of shock self-reformation.

\section{Observations and results}

An inbound crossing of Earth's bow shock is illustrated from MMS observations on 11 January 2016 (Figure 1a). Figure 1b-d shows the relative positions of the satellites in a coordinate system $\mathbf{n}$--$\mathbf{t}_1$--$\mathbf{t}_2$ as defined by Johlander et al. \citep{Johlander2016,Johlander2018}, where $\mathbf{n}$ is the shock normal direction, $\mathbf{t}_2 = \mathbf{n} \times \mathbf{B_u} / |n\times B_u|$, and $\mathbf{t}_1 = \mathbf{t}_2 \times \mathbf{n}$. Here $\mathbf{B_u}$ is the upstream magnetic field vector averaged from 05:37:48 UT to 05:37:58 UT. The shock normal is calculated with a multi-spacecraft timing method \citep{Russell1983,Schwartz1998}. Typical magnetic field peaks around the overshoot are used for the timing method. The obtained value of $\mathbf{n}$ is similar to those from single spacecraft methods and bow shock models \citep{Schwartz1998}. The shock is moving outward at a speed of $\sim105\pm20$ km/s at 05:38:08 UT in the Earth's frame (also given by the multi-spacecraft timing method \citep{Russell1983,Schwartz1998}). In Figure 1b-d, MMS1 is immediately followed by MMS3 and MMS4 in the $\mathbf{n}$ direction during the shock crossing. These three spacecraft (MMS1,3,4), which form a plane nearly parallel to the shock surface, observe roughly the same shock structure (see below). This seems to exclude the possibility of shock rippling. The fourth spacecraft (MMS2), which is separated along $\mathbf{n}$, sees two shock ramps. For MMS1 and MMS2, their separation $\Delta \mathbf{n}(1,2)$ is the largest among the four spacecraft and is about $35$ km ($\sim0.5$ $d_i$, where $d_i$ is the ion inertial length and equals to about 70 km in this case). Corresponding separations $\Delta \mathbf{t}_1(1,2)$ and $\Delta \mathbf{t}_2(1,2)$ are 0.18 $d_i$ and 0.04 $d_i$, respectively, so MMS1 and MMS2 cross the shock front one after another at nearly the same location in the $\mathbf{t}_1$ and $\mathbf{t}_2$ directions. Such a unique configuration of spacecraft provides an opportunity to prove the self-reformation, which will be elaborated below.

Figure 2 shows an overview of the non-stationary shock profiles and corresponding ion velocity distributions observed by MMS1-4. This shock crossing at 05:38:05 UT is associated with an angle between the shock normal and upstream magnetic field $\theta_{Bn}\sim78\pm5^{\circ}$, Alfv\'{e}n Mach number $M_A\sim10.8\pm0.6$, and upstream ion $\beta_i\sim0.3\pm0.02$\renewcommand{\thefootnote}{\roman{footnote}}\footnote{The FPI instrument on MMS is primarily designed to make accurate measurements of the magnetospheric plasma. Measurements of the cold solar wind beam may be less accurate. However, once the plasma starts being heated from the shock foot the plasma measurements by FPI become reliable. To obtain more accurate plasma parameters far upstream of the shock we have used data from Wind \citep{Ogilvie1995} that is situated upstream of MMS at L1.}. The distance between Wind and MMS is about 1.5 million km, and the solar wind speed upstream of the shock is about 400 km/s. We trace the solar wind parameters from MMS back to Wind, which is about an hour's time difference. Because the shock is observed by MMS at about 05:38 UT, we consider the Wind data in the time range from 04:38 UT to 04:39 UT in the above parameter calculation. This is a super-critical, quasi-perpendicular shock with a relatively high ion thermal Mach number, which is in favor of shock front self-reformation \citep{Hada2003,Scholer2003}. In general, the spacecraft except MMS2 observe similar magnetic field profiles and particle distributions. A steep shock ramp with width $<0.3 d_i$ is observed by MMS1, 3 and 4 at about t = 05:38:05.750 UT (vertical dashed line in Figure 2a,e,g). If the shock front were stationary, MMS2 would see the same ramp about 0.3 s after MMS1. However, this is not the case. MMS2 observes a new ramp at the same time (marked by the left dashed line in Figure 2c), which is not yet as large as those seen at other spacecraft. This new ramp could contain two or more peaks due to the nonlinear evolution of the magnetic field during the self-reformation \citep{Scholer2003,Yang2009}, as observed here. A degraded old ramp is also seen at t = 05:38:07.400 UT (marked by the right dashed line in Figure 2c), about 1.7 s after the new ramp, with a lower magnetic field ($<$43 nT). The upstream proton cyclotron time $\Omega_{ci}^{-1}$ is about 1.6 s. The self-reformation period is about 1.7 $\Omega_{ci}^{-1}$ from previous simulations \citep{Hada2003,Lee2005,Yang2009}, which is about 2.7 s in our case. The 1.7 s interval enables MMS2 to observe the evolution of ion kinetic features around the old and new ramps within one self-reformation cycle. We will show that such multiple crossings of shock ramps at MMS2 are caused by the shock self-reformation, i.e., the growing foot ahead of the old ramp, rather than shock surface ripples or the back and forth swings of the shock.

In order to verify that the observed shock front is undergoing self-reformation, we focus on the multiple crossings of shock ramps observed by MMS2 (marked in Figure 3 by dashed lines at 05:38:05.750 UT and 05:38:07.400 UT, respectively). Even though the magnetic field values observed by MMS2 at the two ramps are similar, the electron density $N_e$ (Figure 3b) at the growing foot (i.e., the new ramp) is relatively low compared with that at the old ramp. It is worth noting that the two large amplitude peaks in the magnetic field are not correlated with similar enhancements in the electron density. Simple MHD theory suggests that for a perpendicular shock the magnetic field and density should be compressed by the same ratio. This indicates that the new ramp may not be mature yet. The electron temperature peak (Figure 3c) at the growing foot is lower than that at the old ramp. The new ramp is associated with gyro-reflected ions (Figure 3d) and a decelerated solar wind beam, which leads to an increase in the ion temperature perpendicular to the magnetic field direction (Figure 3c) due to phase mixing \citep{Scholer2004}. The density and temperature profiles should be similar if the multiple crossings are caused by the rippling effect or the back and forth swings of the shock. Obviously, this is not the case, which indicates the possibility of shock self-reformation. Note that the whole time range of Figure 3 is about 6.5 seconds. Cluster measurements can provide at most two moments of data for this range (4 s resolution) and could not distinguish between shock ripples and self-reformation.

For further confirmation of the shock front self-reformation, we examine the ion phase-space distributions (also see the associated animation of Figure 3 online, which shows a 3-D view of ion velocity distributions observed by MMS2 during the shock crossing from 05:37:44 UT to 05:38:30 UT). Figure 3d shows that the incident and reflected ions form a phase-space vortex between the old and new ramps. Such a vortex is different from that observed at a quasi-perpendicular rippling shock; at rippling shocks, the ions are reflected at all the ramps \citep[also see below]{Johlander2018}. The bulk velocity $V_n$ of the incident solar wind ions seen at MMS2 is slowed down (from about $-350$ km/s to $-225$ km/s) and forms a flat profile from the leading edge of the new ramp all the way to the old ramp (Figure 3d). In particular, ion reflection does not occur at the new ramp, and all incident ions are directly transmitted. The ion distributions are distinct from those at MMS1 (Figure 3e) and other two spacecraft as well (Figure 2). The ion distribution features at MMS2 agree well with previous simulations of shock self-reformation \citep{Hada2003,Umeda2006}, which provides a clear evidence for the shock front self-reformation.

Figure 4a shows a schematic profile of the magnetic field and ion phase-space distributions $f(V_n,time)$ for a typical self-reforming shock. At the old ramp (Figure 4d, a high $B$ region), a fraction of incident ions are being reflected. However, at the new ramp (another high $B$ region) all the incident solar wind ions are directly transmitted, and the reflected ions by the old ramp and the incident solar wind ions are clearly separated in the velocity space (Figure 4b). It indicates that the ion reflection is not occurring in this high $B$ region. This velocity distribution is similar to that observed in the low $B$ region between the two ramps (Figure 4c). In contrast, in typical rippling shock observations, ion reflections are observed at all the high $B$ regions \citep{Johlander2018}. These difference reinforce our interpretation of self-reformation for the present case.

\section{Conclusions and discussions}

Based on the above analysis, we conclude that the multiple ramps observed at MMS2 around 05:38 UT on January 11, 2016 during an inbound crossing of Earth's bow shock are caused by the shock front self-reformation along the shock normal, rather than ripples in the shock surface or the back and forth swings of the shock. The multi-point, high-resolution measurements from MMS provide evidence for shock self-reformation, i.e., the phase-space vortex in ion velocity distributions that results in the new shock ramp. In particular, the incident ions observed by MMS2 show a flat velocity profile from the new ramp all the way to the old one. This is because the new ramp has not yet developed to a mature stage during the self-reformation, and is not strong enough to reflect these ions. At the growing foot, the reflected ions by the old ramp and the incident ions are clearly separated. The observed details of the plasma and magnetic field profiles across the shock show that it is the ion reflection that drives the self-reformation of collisionless shocks.

In this letter, the magnetic peaks observed by MMS2 at the new ramp are not correlated with similar enhancements in the electron density. This implies that there may be mechanisms other than the ion accumulation which drives the growth of the new ramp. For example, the decoupling of density and magnetic field compression could imply sub-ion scale physics driven by Weibel instability at high Mach number shocks \citep{Sundberg2017}. Alternatively, since the reformation process is continually dynamic, perhaps this signature can be understood as a transition on e.g. a whistler time scale \citep{Krasnoselskikh2002}. Numerical studies also suggest that, modified two-stream instability \citep{Scholer2003,Scholer2004} may cause multiple magnetic field peaks. Furthermore, observations of high Mach number shocks evidence that shock parameters can affect the self-reformation period \citep{Sulaiman2015}.

Insights gained from this study will help understand the particle acceleration mechanism and microstructures \citep{Yang2009,Matsukiyo2011} associated with nonstationary collisionless shocks as well as radio emissions \citep{Liu2009,Morosan2019} and energy dissipation \citep{Parks2012,Yang2014} throughout the universe.

\acknowledgments

We are grateful to the valuable suggestions from A. Vaivads from Swedish Institute of Space Physics, and S. J. Schwartz from University of Colorado, Boulder. Z.W.Y. appreciates helpful discussions with R. S. Wang from USTC, B. B. Tang and W. Y. Li from NSSC. We acknowledge the use of data from MMS (https://lasp.colorado.edu/mms/sdc/), the database on NASA CDAWeb, and the clweb maintained by E. Penou. This work is jointly supported by the NSFC under Grants 41574140, 41674168 and 41774179, the Specialized Research Fund for State Key Laboratories of China, Youth Innovation Promotion Association of the CAS (2017188), the Open Research Program Key laboratory of Polar Science, MNR (KP202005), NSF of Beijing Municipality (1192018), Beijing Outstanding Talent Training Foundation (2017000097607G049), Beijing Municipal Science and Technology Commission (grant No. Z191100004319001 and Z191100004319003), and the Strategic Priority Research Program of CAS (Grant No. XDA14040404).\\


%\vspace{5mm}


%\bibliography{sample63}{}
\bibliography{references}{}

\begin{thebibliography}{}
\expandafter\ifx\csname natexlab\endcsname\relax\def\natexlab#1{#1}\fi
\providecommand{\url}[1]{\href{#1}{#1}}
\providecommand{\dodoi}[1]{doi:~\href{http://doi.org/#1}{\nolinkurl{#1}}}
\providecommand{\doeprint}[1]{\href{http://ascl.net/#1}{\nolinkurl{http://ascl.net/#1}}}
\providecommand{\doarXiv}[1]{\href{https://arxiv.org/abs/#1}{\nolinkurl{https://arxiv.org/abs/#1}}}

\bibitem[{{Bale} {et~al.}(2005){Bale}, {Balikhin}, {Horbury}, {Krasnoselskikh},
  {Kucharek}, {M{\"o}bius}, {Walker}, {Balogh}, {Burgess}, {Lemb{\`e}ge},
  {Lucek}, {Scholer}, {Schwartz7 10}, \& {Thomsen}}]{Bale2005}
{Bale}, S.~D., {Balikhin}, M.~A., {Horbury}, T.~S., {et~al.} 2005, Space Sci.
  Rev., 118, 161, \dodoi{10.1007/s11214-005-3827-0}

\bibitem[{{Blandford} \& {Eichler}(1987)}]{Blandford1987}
{Blandford}, R., \& {Eichler}, D. 1987, Phys. Rep., 154, 1,
  \dodoi{10.1016/0370-1573(87)90134-7}

\bibitem[{{Burch} {et~al.}(2016){Burch}, {Moore}, {Torbert}, \&
  {Giles}}]{Burch2016}
{Burch}, J.~L., {Moore}, T.~E., {Torbert}, R.~B., \& {Giles}, B.~L. 2016, Space
  Sci. Rev., \dodoi{10.1007/s11214-015-0164-9}

\bibitem[{{Burgess} {et~al.}(2005){Burgess}, {Lucek}, {Scholer}, {Bale},
  {Balikhin}, {Balogh}, {Horbury}, {Krasnoselskikh}, {Kucharek}, {Lemb{\`e}ge},
  {M{\"o}bius}, {Schwartz}, {Thomsen}, \& {Walker}}]{Burgess2005}
{Burgess}, D., {Lucek}, E.~A., {Scholer}, M., {et~al.} 2005, Space Sci. Rev.,
  118, 205, \dodoi{10.1007/s11214-005-3832-3}

\bibitem[{{Caprioli} {et~al.}(2015){Caprioli}, {Pop}, \&
  {Spitkovsky}}]{Caprioli2015}
{Caprioli}, D., {Pop}, A.-R., \& {Spitkovsky}, A. 2015, Astrophys. J. Lett.,
  798, L28, \dodoi{10.1088/2041-8205/798/2/L28}

\bibitem[{{Chapman} {et~al.}(2005){Chapman}, {Lee}, \& {Dendy}}]{Chapman2005}
{Chapman}, S.~C., {Lee}, R.~E., \& {Dendy}, R.~O. 2005, Space Sci. Rev., 121,
  5, \dodoi{10.1007/s11214-006-4481-x}

\bibitem[{{Gingell} {et~al.}(2017){Gingell}, {Schwartz}, {Burgess},
  {Johlander}, {Russell}, {Burch}, {Ergun}, {Fusellier}, {Gershman}, {Giles},
  {Goodrich}, {Khotyaintsev}, {Lavraud}, {Lindqvist}, {Strangeway}, {Trattner},
  {Torbert}, {Wei}, \& {Wilder}}]{Gingell2017}
{Gingell}, B., {Schwartz}, S.~J., {Burgess}, D., {et~al.} 2017, J. Geophys.
  Res., 122, 11,003, \dodoi{10.1002/2017JA024538}

\bibitem[{{Guo} {et~al.}(2012){Guo}, {Li}, {Li}, {Giacalone}, {Jokipii}, \&
  {Li}}]{Guo2012}
{Guo}, F., {Li}, S., {Li}, H., {et~al.} 2012, \apj, 747, 98,
  \dodoi{10.1088/0004-637X/747/2/98}

\bibitem[{{Hada} {et~al.}(2003){Hada}, {Oonishi}, {Lemb{\`e}ge}, \&
  {Savoini}}]{Hada2003}
{Hada}, T., {Oonishi}, M., {Lemb{\`e}ge}, B., \& {Savoini}, P. 2003, J.
  Geophys. Res., 108, 1233, \dodoi{10.1029/2002JA009339}

\bibitem[{{Hao} {et~al.}(2017){Hao}, {Gao}, {Lu}, {Huang}, {Wang}, \&
  {Wang}}]{Hao2017}
{Hao}, Y.~F., {Gao}, X.~L., {Lu}, Q.~M., {et~al.} 2017, J. Geophys. Res., 122,
  6385, \dodoi{10.1002/2017JA024234}

\bibitem[{{Horbury} {et~al.}(2001){Horbury}, {Cargill}, {Lucek}, {Balogh},
  {Dunlop}, {Oddy}, {Carr}, {Brown}, {Szabo}, \& {Fornacon}}]{Horbury2001}
{Horbury}, T.~S., {Cargill}, P.~J., {Lucek}, E.~A., {et~al.} 2001, Ann.
  Geophys., 19, 1399, \dodoi{10.5194/angeo-19-1399-2001}

\bibitem[{{Johlander} {et~al.}(2018){Johlander}, {Vaivads}, {Khotyaintsev},
  {Gingell}, {Schwartz}, {Giles}, {Torbert}, \& {Russell}}]{Johlander2018}
{Johlander}, A., {Vaivads}, A., {Khotyaintsev}, Y.~V., {et~al.} 2018, Plasma
  Phys. Control. Fusion, 60, 125006, \dodoi{10.1088/1361-6587/aae920}

\bibitem[{{Johlander} {et~al.}(2016){Johlander}, {Schwartz}, {Vaivads},
  {Khotyaintsev}, {Gingell}, {Peng}, {Markidis}, {Lindqvist}, {Ergun},
  {Marklund}, {Plaschke}, {Magnes}, {Strangeway}, {Russell}, {Wei}, {Torbert},
  {Paterson}, {Gershman}, {Dorelli}, {Avanov}, {Lavraud}, {Saito}, {Giles},
  {Pollock}, \& {Burch}}]{Johlander2016}
{Johlander}, A., {Schwartz}, S.~J., {Vaivads}, A., {et~al.} 2016, Phys. Rev.
  Lett., 117, 165101, \dodoi{10.1103/PhysRevLett.117.165101}

\bibitem[{{Krasnoselskikh} {et~al.}(2002){Krasnoselskikh}, {Lemb{\`e}ge},
  {Savoini}, \& {Lobzin}}]{Krasnoselskikh2002}
{Krasnoselskikh}, V.~V., {Lemb{\`e}ge}, B., {Savoini}, P., \& {Lobzin}, V.~V.
  2002, Phys. Plasmas, 9, 1192, \dodoi{10.1063/1.1457465}

\bibitem[{{Lee} {et~al.}(2005){Lee}, {Chapman}, \& {Dendy}}]{Lee2005}
{Lee}, R.~E., {Chapman}, S.~C., \& {Dendy}, R.~O. 2005, Phys. Plasmas, 12,
  012901, \dodoi{10.1063/1.1812536}

\bibitem[{{Lemb{\`e}ge} \& {Dawson}(1987)}]{Lembege1987}
{Lemb{\`e}ge}, B., \& {Dawson}, J.~M. 1987, Phys. Fluids, 30, 1767,
  \dodoi{10.1063/1.866191}

\bibitem[{{Lembege} \& {Savoini}(1992)}]{Lembege1992}
{Lembege}, B., \& {Savoini}, P. 1992, Phys. Fluids B, 4, 3533,
  \dodoi{10.1063/1.860361}

\bibitem[{{Lemb{\`e}ge} {et~al.}(2009){Lemb{\`e}ge}, {Savoini}, {Hellinger}, \&
  {Tr{\'a}vn{\'\i}{\v{c}}ek}}]{Lembege2009}
{Lemb{\`e}ge}, B., {Savoini}, P., {Hellinger}, P., \&
  {Tr{\'a}vn{\'\i}{\v{c}}ek}, P.~M. 2009, J. Geophys. Res., 114, A03217,
  \dodoi{10.1029/2008JA013618}

\bibitem[{{Lemb{\`e}ge} {et~al.}(2004){Lemb{\`e}ge}, {Giacalone}, {Scholer},
  {Hada}, {Hoshino}, {Krasnoselskikh}, {Kucharek}, {Savoini}, \&
  {Terasawa}}]{Lembege2004}
{Lemb{\`e}ge}, B., {Giacalone}, J., {Scholer}, M., {et~al.} 2004, Space Sci.
  Rev., 110, 161, \dodoi{10.1023/B:SPAC.0000023372.12232.b7}

\bibitem[{{Liu} {et~al.}(2009){Liu}, {Luhmann}, {Bale}, \& {Lin}}]{Liu2009}
{Liu}, Y., {Luhmann}, J.~G., {Bale}, S.~D., \& {Lin}, R.~P. 2009, \apjl, 691,
  L151, \dodoi{10.1088/0004-637X/691/2/L151}

\bibitem[{{Liu} {et~al.}(2019){Liu}, {Zhu}, \& {Zhao}}]{Liu2019}
{Liu}, Y.~D., {Zhu}, B., \& {Zhao}, X. 2019, \apj, 871, 8,
  \dodoi{10.3847/1538-4357/aaf425}

\bibitem[{{Lobzin} {et~al.}(2007){Lobzin}, {Krasnoselskikh}, {Bosqued},
  {Pin{\c{c}}on}, {Schwartz}, \& {Dunlop}}]{Lobzin2007}
{Lobzin}, V.~V., {Krasnoselskikh}, V.~V., {Bosqued}, J.~M., {et~al.} 2007,
  Geophys. Res. Lett., 34, L05107, \dodoi{10.1029/2006GL029095}

\bibitem[{{Matsukiyo} {et~al.}(2011){Matsukiyo}, {Ohira}, {Yamazaki}, \&
  {Umeda}}]{Matsukiyo2011}
{Matsukiyo}, S., {Ohira}, Y., {Yamazaki}, R., \& {Umeda}, T. 2011, \apj, 742,
  47, \dodoi{10.1088/0004-637X/742/1/47}

\bibitem[{{Mazelle} {et~al.}(2010){Mazelle}, {Lemb{\`e}ge}, {Morgenthaler},
  {Meziane}, {Horbury}, {G{\'e}not}, {Lucek}, \& {Dandouras}}]{Mazelle2010}
{Mazelle}, C., {Lemb{\`e}ge}, B., {Morgenthaler}, A., {et~al.} 2010, AIP Conf.
  Proc., 1216, 471, \dodoi{10.1063/1.3395905}

\bibitem[{{Morosan} {et~al.}(2019){Morosan}, {Carley}, {Hayes}, {Murray},
  {Zucca}, {Fallows}, {McCauley}, {Kilpua}, {Mann}, {Vocks}, \&
  {Gallagher}}]{Morosan2019}
{Morosan}, D.~E., {Carley}, E.~P., {Hayes}, L.~A., {et~al.} 2019, Nature
  Astronomy, 252, \dodoi{10.1038/s41550-019-0689-z}

\bibitem[{{Ogilvie} {et~al.}(1995){Ogilvie}, {Chornay}, {Fritzenreiter},
  {Hunsaker}, {Keller}, {Lobell}, {Miller}, {Scudder}, {Sittler}, {Torbert},
  {Bodet}, {Needell}, {Lazarus}, {Steinberg}, {Tappan}, {Mavretic}, \&
  {Gergin}}]{Ogilvie1995}
{Ogilvie}, K.~W., {Chornay}, D.~J., {Fritzenreiter}, R.~J., {et~al.} 1995,
  Space Sci. Rev., 71, 55, \dodoi{10.1007/BF00751326}

\bibitem[{{Parks} {et~al.}(2012){Parks}, {Lee}, {McCarthy}, {Goldstein}, {Fu},
  {Cao}, {Canu}, {Lin}, {Wilber}, {Dandouras}, {R{\'e}me}, \&
  {Fazakerley}}]{Parks2012}
{Parks}, G.~K., {Lee}, E., {McCarthy}, M., {et~al.} 2012, Phys. Rev. Lett.,
  108, 061102, \dodoi{10.1103/PhysRevLett.108.061102}

\bibitem[{{Plotnikov} \& {Sironi}(2019)}]{Plotnikov2019}
{Plotnikov}, I., \& {Sironi}, L. 2019, \mnras, 490, 156,
  \dodoi{10.1093/mnras/stz2653}

\bibitem[{{Pollock} {et~al.}(2016){Pollock}, {Moore}, {Jacques}, {Burch},
  {Gliese}, {Saito}, {Omoto}, {Avanov}, {Barrie}, {Coffey}, {Dorelli},
  {Gershman}, {Giles}, {Rosnack}, {Salo}, {Yokota}, {Adrian}, {Aoustin},
  {Auletti}, {Aung}, {Bigio}, {Cao}, {Chandler}, {Chornay}, {Christian},
  {Clark}, {Collinson}, {Corris}, {De Los Santos}, {Devlin}, {Diaz},
  {Dickerson}, {Dickson}, {Diekmann}, {Diggs}, {Duncan}, {Figueroa-Vinas},
  {Firman}, {Freeman}, {Galassi}, {Garcia}, {Goodhart}, {Guererro}, {Hageman},
  {Hanley}, {Hemminger}, {Holland}, {Hutchins}, {James}, {Jones}, {Kreisler},
  {Kujawski}, {Lavu}, {Lobell}, {LeCompte}, {Lukemire}, {MacDonald}, {Mariano},
  {Mukai}, {Narayanan}, {Nguyan}, {Onizuka}, {Paterson}, {Persyn}, {Piepgrass},
  {Cheney}, {Rager}, {Raghuram}, {Ramil}, {Reichenthal}, {Rodriguez},
  {Rouzaud}, {Rucker}, {Saito}, {Samara}, {Sauvaud}, {Schuster}, {Shappirio},
  {Shelton}, {Sher}, {Smith}, {Smith}, {Smith}, {Steinfeld}, {Szymkiewicz},
  {Tanimoto}, {Taylor}, {Tucker}, {Tull}, {Uhl}, {Vloet}, {Walpole}, {Weidner},
  {White}, {Winkert}, {Yeh}, \& {Zeuch}}]{Pollock2016}
{Pollock}, C., {Moore}, T., {Jacques}, A., {et~al.} 2016, Space Sci. Rev., 199,
  331, \dodoi{10.1007/s11214-016-0245-4}

\bibitem[{{Richardson} {et~al.}(2008){Richardson}, {Kasper}, {Wang}, {Belcher},
  \& {Lazarus}}]{Richardson2008}
{Richardson}, J.~D., {Kasper}, J.~C., {Wang}, C., {Belcher}, J.~W., \&
  {Lazarus}, A.~J. 2008, Nature, 454, 63, \dodoi{10.1038/nature07024}

\bibitem[{{Russell} {et~al.}(1983){Russell}, {Mellott}, {Smith}, \&
  {King}}]{Russell1983}
{Russell}, C.~T., {Mellott}, M.~M., {Smith}, E.~J., \& {King}, J.~H. 1983,
  \jgr, 88, 4739, \dodoi{10.1029/JA088iA06p04739}

\bibitem[{{Schaeffer} {et~al.}(2019){Schaeffer}, {Fox}, {Follett}, {Fiksel},
  {Li}, {Matteucci}, {Bhattacharjee}, \& {Germaschewski}}]{Schaeffer2019}
{Schaeffer}, D.~B., {Fox}, W., {Follett}, R.~K., {et~al.} 2019, Phys. Rev.
  Lett., 122, 245001, \dodoi{10.1103/PhysRevLett.122.245001}

\bibitem[{{Scholer} \& {Matsukiyo}(2004)}]{Scholer2004}
{Scholer}, M., \& {Matsukiyo}, S. 2004, Ann. Geophys., 22, 2345,
  \dodoi{10.5194/angeo-22-2345-2004}

\bibitem[{{Scholer} {et~al.}(2003){Scholer}, {Shinohara}, \&
  {Matsukiyo}}]{Scholer2003}
{Scholer}, M., {Shinohara}, I., \& {Matsukiyo}, S. 2003, J. Geophys. Res., 108,
  1014, \dodoi{10.1029/2002JA009515}

\bibitem[{{Schwartz}(1998)}]{Schwartz1998}
{Schwartz}, S.~J. 1998, ISSI Sci. Rep. Ser., 1, 249

\bibitem[{{Sulaiman} {et~al.}(2015){Sulaiman}, {Masters}, {Dougherty},
  {Burgess}, {Fujimoto}, \& {Hospodarsky}}]{Sulaiman2015}
{Sulaiman}, A.~H., {Masters}, A., {Dougherty}, M.~K., {et~al.} 2015, \prl, 115,
  125001, \dodoi{10.1103/PhysRevLett.115.125001}

\bibitem[{{Sundberg} {et~al.}(2017){Sundberg}, {Burgess}, {Scholer}, {Masters},
  \& {Sulaiman}}]{Sundberg2017}
{Sundberg}, T., {Burgess}, D., {Scholer}, M., {Masters}, A., \& {Sulaiman},
  A.~H. 2017, Astrophys. J. Lett., 836, 1, \dodoi{10.3847/2041-8213/836/1/L4}

\bibitem[{{Treumann}(2009)}]{Treumann2009}
{Treumann}, R.~A. 2009, Astron. Astrophys. Rev., 17, 409,
  \dodoi{10.1007/s00159-009-0024-2}

\bibitem[{{Umeda} \& {Daicho}(2018)}]{Umeda2018}
{Umeda}, T., \& {Daicho}, Y. 2018, Ann. Geophys., 36, 1047,
  \dodoi{10.5194/angeo-36-1047-2018}

\bibitem[{{Umeda} \& {Yamazaki}(2006)}]{Umeda2006}
{Umeda}, T., \& {Yamazaki}, R. 2006, Earth, Planets, and Space, 58, e41,
  \dodoi{10.1186/BF03352617}

\bibitem[{{Wilson} {et~al.}(2019){Wilson}, {Chen}, {Wang}, {Schwartz},
  {Turner}, {Stevens}, {Kasper}, {Osmane}, {Caprioli}, {Bale}, {Pulupa},
  {Salem}, \& {Goodrich}}]{Wilson2019}
{Wilson}, Lynn~B., I., {Chen}, L.-J., {Wang}, S., {et~al.} 2019, Astrophys. J.
  Supp. Ser., 243, 8, \dodoi{10.3847/1538-4365/ab22bd}

\bibitem[{{Wilson} {et~al.}(2020){Wilson}, {Chen}, {Wang}, {Schwartz},
  {Turner}, {Stevens}, {Kasper}, {Osmane}, {Caprioli}, {Bale}, {Pulupa},
  {Salem}, \& {Goodrich}}]{Wilson2020}
---. 2020, \apj, 893, 22, \dodoi{10.3847/1538-4357/ab7d39}

\bibitem[{{Yang} {et~al.}(2014){Yang}, {Liu}, {Parks}, {Wu}, {Huang}, {Shi},
  {Wang}, \& {Hu}}]{Yang2014}
{Yang}, Z., {Liu}, Y.~D., {Parks}, G.~K., {et~al.} 2014, Astrophys. J. Lett.,
  793, L11, \dodoi{10.1088/2041-8205/793/1/L11}

\bibitem[{{Yang} {et~al.}(2018){Yang}, {Lu}, {Liu}, \& {Wang}}]{Yang2018}
{Yang}, Z., {Lu}, Q., {Liu}, Y.~D., \& {Wang}, R. 2018, Astrophys. J., 857, 36,
  \dodoi{10.3847/1538-4357/aab714}

\bibitem[{{Yang} {et~al.}(2009){Yang}, {Lu}, {Lemb{\`e}ge}, \&
  {Wang}}]{Yang2009}
{Yang}, Z.~W., {Lu}, Q.~M., {Lemb{\`e}ge}, B., \& {Wang}, S. 2009, J. Geophys.
  Res., 114, A03111, \dodoi{10.1029/2008JA013785}

\bibitem[{{Zank}(2015)}]{Zank2015a}
{Zank}, G.~P. 2015, \araa, 53, 449, \dodoi{10.1146/annurev-astro-082214-122254}

\bibitem[{{Zank} {et~al.}(2015){Zank}, {Hunana}, {Mostafavi}, {Le Roux}, {Li},
  {Webb}, {Khabarova}, {Cummings}, {Stone}, \& {Decker}}]{Zank2015b}
{Zank}, G.~P., {Hunana}, P., {Mostafavi}, P., {et~al.} 2015, \apj, 814, 137,
  \dodoi{10.1088/0004-637X/814/2/137}

\end{thebibliography}

\clearpage

\begin{figure}
\epsscale{1.0}
\plotone{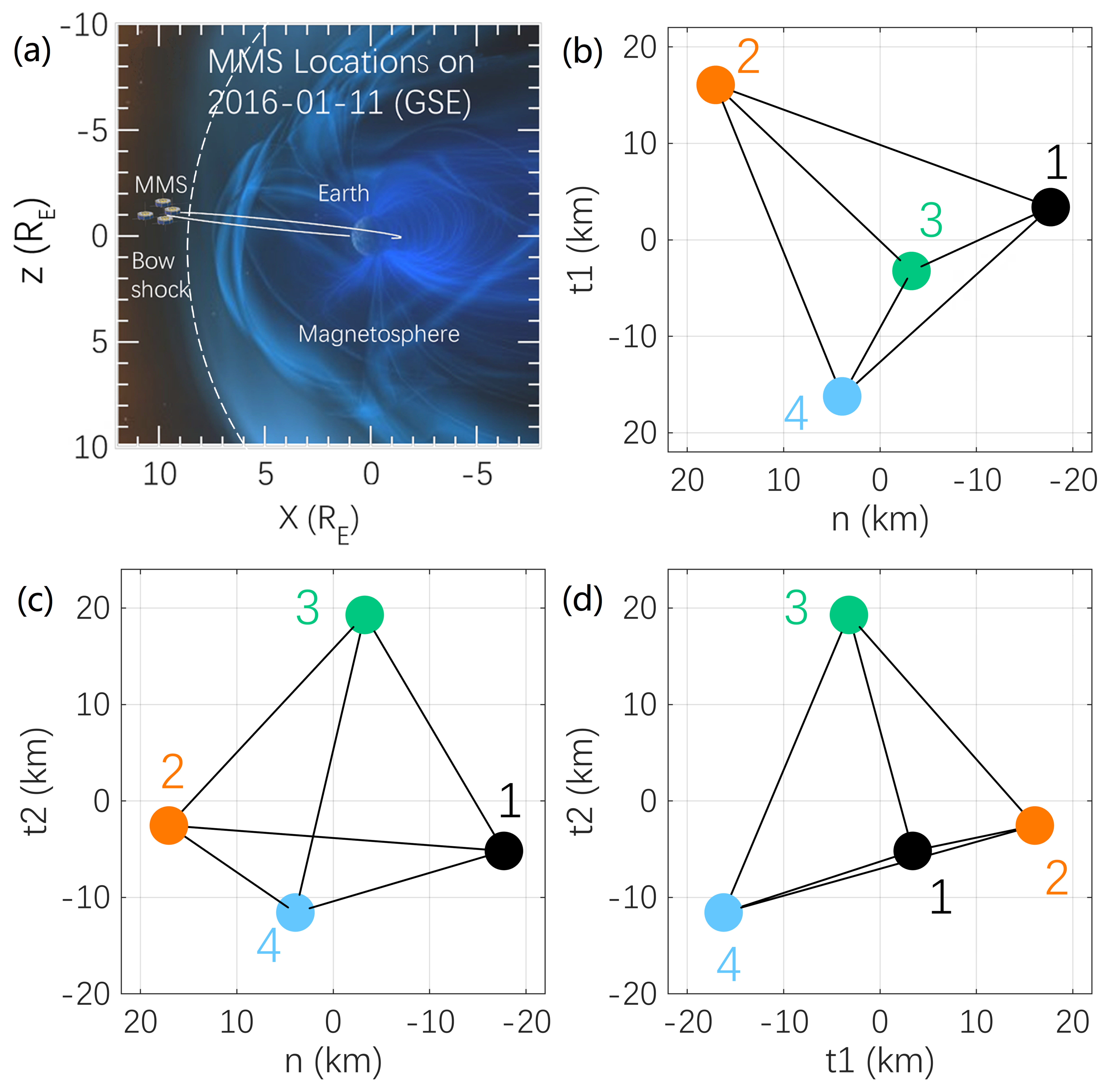}
\caption{\label{fig:fig1}Geometry of MMS with respect to Earth's bow shock. (a) The locations of the four MMS spacecraft during an inbound bow shock crossing on January 11, 2016, 05:38:00 UT. The white solid curve is the spacecraft orbit on that day. An artistic illustration (modified from https://www.nasa.gov/) is taken as the background to represent the relative locations of Earth's bow shock (white dashed curve), magnetosphere and the Earth. (b-d) The configuration of the four spacecraft in $\mathbf{n}$--$\mathbf{t}_1$--$\mathbf{t}_2$ coordinates. Black, orange, green and blue dots indicate the four spacecraft MMS1-4, respectively. The separation between MMS1 and MMS2 in the n direction (which can observe the shock transition one by one at a time scale of the ion cyclotron time) provides a good opportunity to investigate the shock front self-reformation process.}
\end{figure}

\clearpage

\begin{figure}
\epsscale{1.1}
\plotone{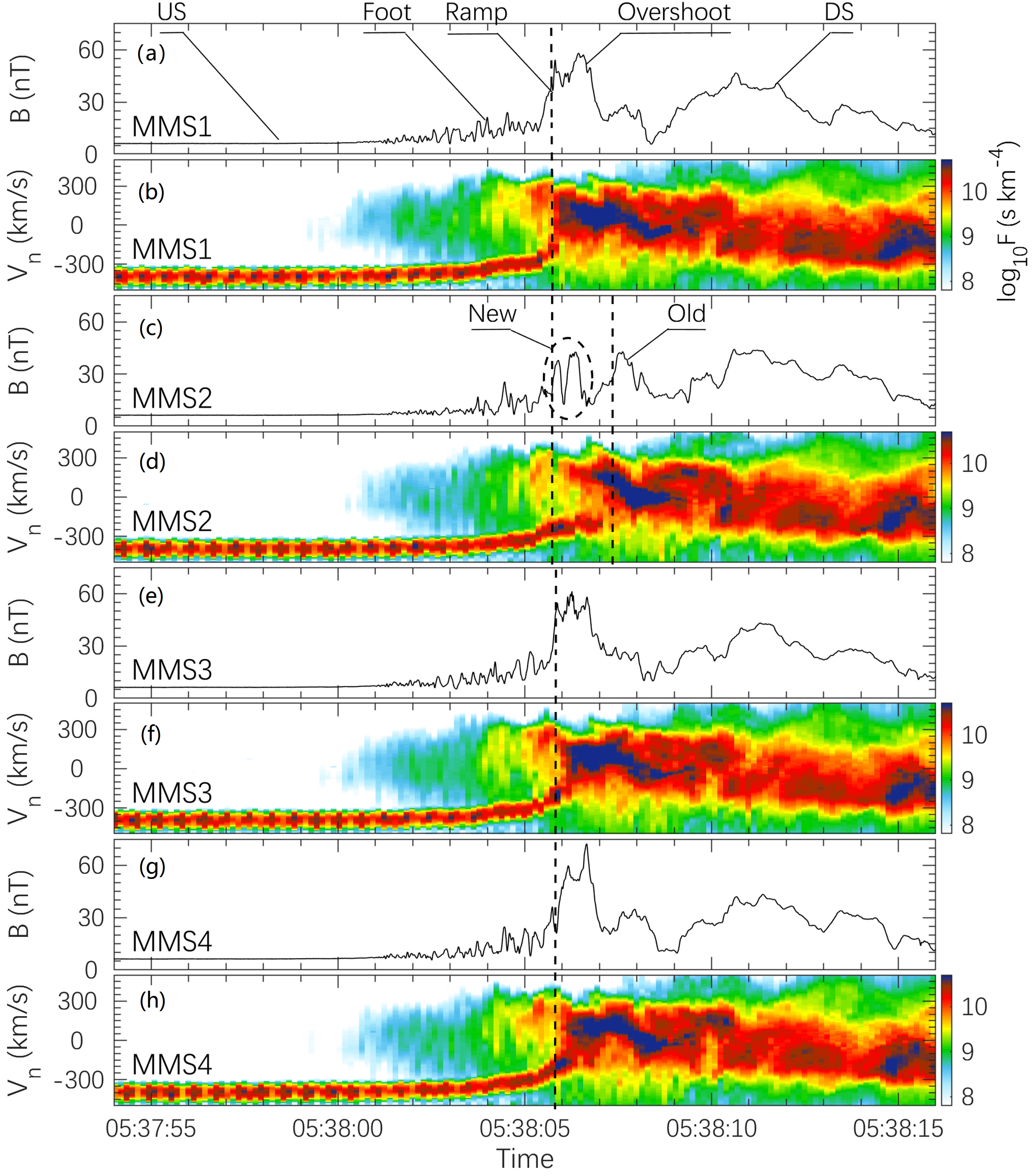}
\caption{\label{fig:fig2}Overview of the Earth's bow shock crossing event on January 11, 2016 by MMS. (a, c, e, and g) Four-spacecraft observations of magnetic field. (b, d, f, and h) Ion velocity distributions with the same colorbar range. The magnetic field profiles observed by MMS1, 3, and 4 are quite similar. MMS2 observes multiple crossings of shock ramps (the old and new ramps are marked by vertical dashed lines). The magnetic field peaks observed at MMS2 are lower than those at MMS1, 3, and 4. The locations of the upstream (US), shock foot, ramp (marked by black dashed lines), overshoot and downstream (DS) are labeled at the top of the figure.}
\end{figure}

\clearpage

\begin{figure}
\epsscale{0.9}
\plotone{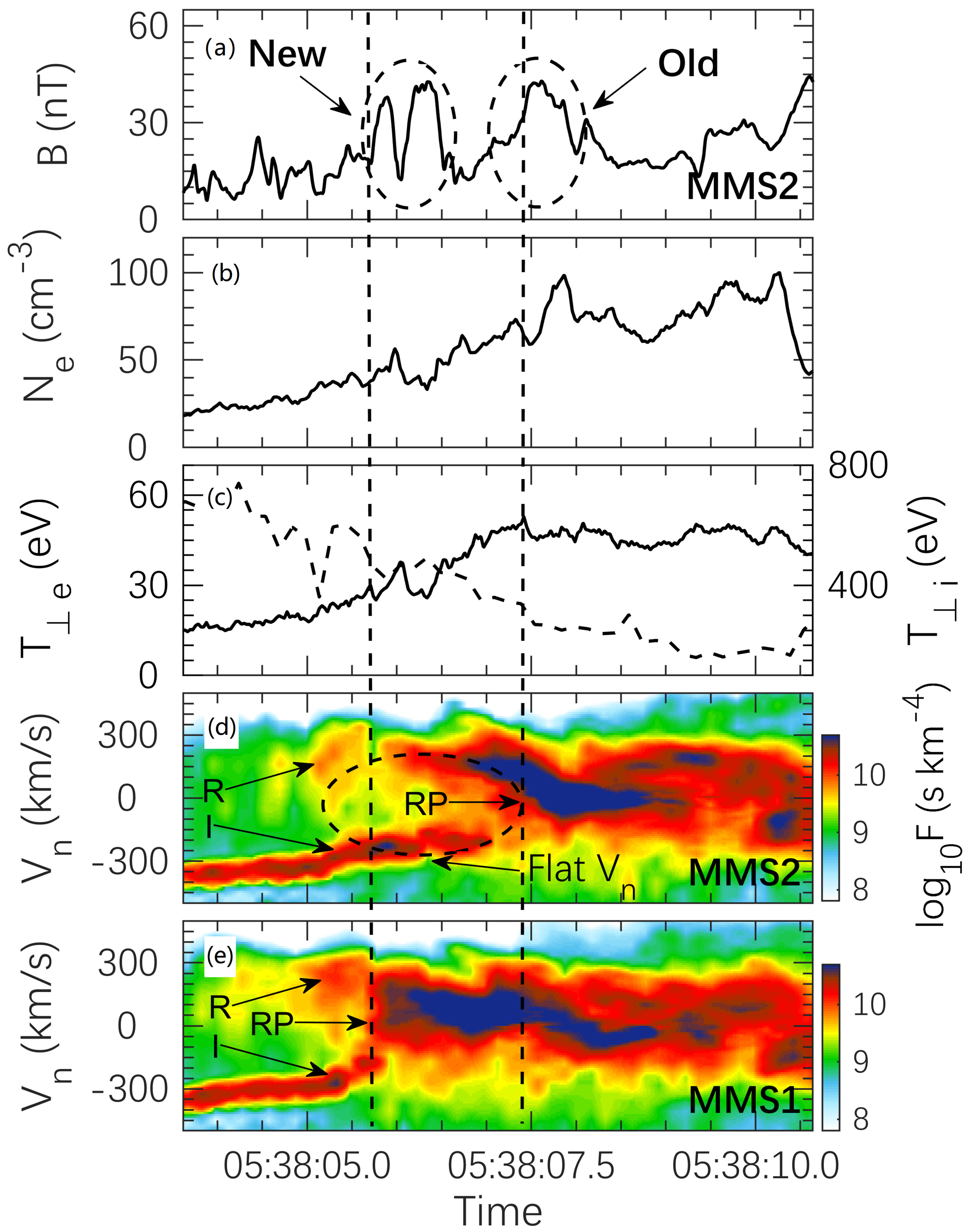}
\caption{\label{fig:fig3}Expanded view of the shock crossings at MMS2. (a) Magnetic field $B$. The old and new ramps are marked by dashed ellipses. (b) Electron density. (c) Electron (solid) and ion (dashed) perpendicular temperature. (d) Phase-space ($time-V_\mathbf{n}$) distribution of ions in log scale ($log_{10}F_{MMS2}$) observed by MMS2. The ion vortex associated with shock self-reformation is marked by the dashed ellipse. Reflected and incident ions are indicated by ``R" and ``I", respectively. The reflection point of ions at the shock ramp is marked by ``RP", and the ``Flat $V_n$" in panel (d) indicates that the solar wind ions directly transmitted the new ramp. (e) Similar to (d) but for MMS1 with the same colorbar range. An animation of panel (d) (a 3-D view of ion velocity distributions from MMS2) is available, which begins at 05:37:44 UT and ends at 05:38:30 UT.}
%\\(An animation of this figure is available.)
\end{figure}

\clearpage

\begin{figure}
\epsscale{1.0}
\plotone{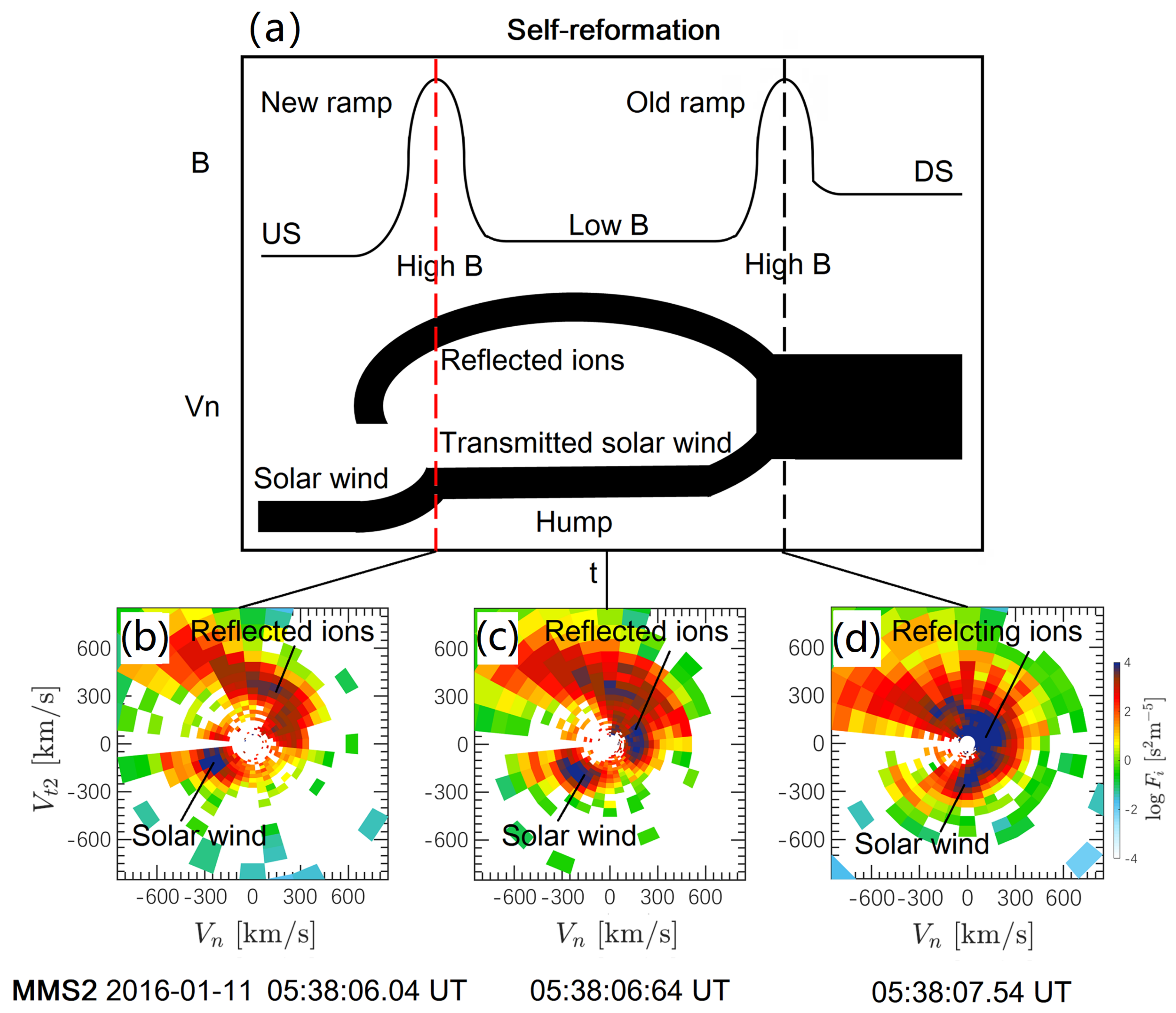}
\caption{\label{fig:fig4}Main features of self-reformation observations. (a) Schematic profiles of the magnetic field and ion phase-space distributions $f(V_n,time)$ for shock self-reformation. (b-d) Ion velocity distributions ($V_n-V_{t2}$) observed in the shock reformation case corresponding to the new and old ramps and the region in between.}
\end{figure}

\clearpage



\end{document}